\documentstyle[referee,epsf]{mn}
\input psfig.tex
\def\lsim{\lower.5ex\hbox{$\; \buildrel < \over \sim \;$}}
\def\gsim{\lower.5ex\hbox{$\; \buildrel > \over \sim \;$}}
\newif\ifAMStwofonts
\ifoldfss
  \ifCUPmtlplainloaded \else
    \NewTextAlphabet{textbfit} {cmbxti10} {}
    \NewTextAlphabet{textbfss} {cmssbx10} {}
    \NewMathAlphabet{mathbfit} {cmbxti10} {} 
    \NewMathAlphabet{mathbfss} {cmssbx10} {} 
  \fi
  \ifAMStwofonts
    \ifCUPmtlplainloaded \else
      \NewSymbolFont{upmath} {eurm10}
      \NewSymbolFont{AMSa} {msam10}
      \NewMathSymbol{\upi}     {0}{upmath}{19}
      \NewMathSymbol{\umu}     {0}{upmath}{16}
      \NewMathSymbol{\upartial}{0}{upmath}{40}
      \NewMathSymbol{\leqslant}{3}{AMSa}{36}
      \NewMathSymbol{\geqslant}{3}{AMSa}{3E}

    \fi
  \fi
\fi 
\ifnfssone
  \newmathalphabet{\mathit}
  \addtoversion{normal}{\mathit}{cmr}{m}{it}
  \addtoversion{bold}{\mathit}{cmr}{bx}{it}
  \newmathalphabet{\mathbfit} 
  \addtoversion{normal}{\mathbfit}{cmr}{bx}{it}
  \addtoversion{bold}{\mathbfit}{cmr}{bx}{it}
  \newmathalphabet{\mathbfss} 
  \addtoversion{normal}{\mathbfss}{cmss}{bx}{n}
  \addtoversion{bold}{\mathbfss}{cmss}{bx}{n}
  \ifAMStwofonts
    \ifCUPmtlplainloaded \else
      \UseAMStwoboldmath
      \makeatletter
      \new@mathgroup\upmath@group
      \define@mathgroup\mv@normal\upmath@group{eur}{m}{n}
      \define@mathgroup\mv@bold\upmath@group{eur}{b}{n}
      \edef\UPM{\hexnumber\upmath@group}
      \new@mathgroup\amsa@group
      \define@mathgroup\mv@normal\amsa@group{msa}{m}{n}
      \define@mathgroup\mv@bold\amsa@group{msa}{m}{n}
      \edef\AMSa{\hexnumber\amsa@group}
      \makeatother
      \mathchardef\upi="0\UPM19
      \mathchardef\umu="0\UPM16
      \mathchardef\upartial="0\UPM40
      \mathchardef\leqslant="3\AMSa36
      \mathchardef\geqslant="3\AMSa3E
    \fi
  \fi
\fi 

\ifnfsstwo
  \DeclareMathAlphabet{\mathbfit}{OT1}{cmr}{bx}{it}
  \SetMathAlphabet\mathbfit{bold}{OT1}{cmr}{bx}{it}
  \DeclareMathAlphabet{\mathbfss}{OT1}{cmss}{bx}{n}
  \SetMathAlphabet\mathbfss{bold}{OT1}{cmss}{bx}{n}
  \ifAMStwofonts
    \ifCUPmtlplainloaded \else
      \SetSymbolFont{UPM}{bold}{U}{eur}{b}{n}
      \DeclareSymbolFont{AMSa}{U}{msa}{m}{n}
      \DeclareMathSymbol{\upi}{0}{UPM}{"19}
      \DeclareMathSymbol{\umu}{0}{UPM}{"16}
      \DeclareMathSymbol{\upartial}{0}{UPM}{"40}
      \DeclareMathSymbol{\leqslant}{3}{AMSa}{"36}
      \DeclareMathSymbol{\geqslant}{3}{AMSa}{"3E}
    \fi
  \fi
\fi 

\ifCUPmtlplainloaded \else
  \ifAMStwofonts \else 
    \def\upi{\pi}
    \def\umu{\mu}
    \def\upartial{\partial}
  \fi
\fi
\def\lsim{\lower.5ex\hbox{$\; \buildrel < \over \sim \;$}}
\def\gsim{\lower.5ex\hbox{$\; \buildrel > \over \sim \;$}}

\title{Molecular Hydrogen Formation During Dense Interstellar Cloud Collapse}

\author[Kinsuk Acharyya, Sandip K.\ Chakrabarti and Sonali Chakrabarti]
{Kinsuk Acharyya$^1$, Sandip K.\ Chakrabarti$^{2,1}$ and Sonali Chakrabarti$^{3,1}$\\
$^1$ Centre for Space Physics, Chalantika 43, Garia Station Rd., Kolkata, 700084, India\\
$^2$ S. N. Bose National Center for Basic Sciences, JD-Block, Salt Lake, Kolkata, 700098, India\\
$^3$  Maharaja Manindra Chandra College, 20 Bhupen Bose Avenue, Kolkata 700003\\  }

\begin{document}

\maketitle

\begin{abstract}
We study the evolution of molecular hydrogen on the grain surfaces and in the gas phase using both the 
rate equation (which tracks the average number of molecules) and the master equation 
(which tracks the expectation values of molecules). We show that above 
a certain critical accretion rate of $H$ on the grains, the results from these two methods become identical.
We used this result to follow the collapse of a dense interstellar cloud and studied the 
formation of molecular hydrogen for two different temperatures (T=10K and 12K) and two different
masses ($1M_\odot$ and $10M_\odot$) of the cloud when olivine grains were
used. Since at higher temperatures, the recombination is very small for these grains,
we also studied a similar hydrodynamic processes at higher temperatures (T=20K and 25K) when amorphous carbon
grains were used. We find that generally, for olivine grains, more than $90$\% $H$ 
is converted to $H_2$  within $\sim 10^{5-7}$yr whereas for amorphous grains it takes
$\sim 10^{6-7}$yr.  $H_2$ formed in this manner can be adequate to produce the observed complex 
molecules.
\end{abstract}

\noindent SUBMITTED FOR PUBLICATION IN MNRAS


\section{Introduction}

Over the last few years several works have been carried out to investigate the formation of 
complex molecules in cool interstellar clouds (Prasad and Huntress 1980a, 1980b; 
Leung, Herbst and H\"ubner 1984; Hasegawa and Herbst 1993, Hasegawa, Herbst and Leung 1992). 
One of the stumbling blocks has been
to identify mechanisms to produce $H_2$ molecules. It is has been realized that 
purely gas phase reactions are so improbable that one needs to invoke
the grain chemistry (e.g., Gould and Salpeter 1963; Hollenbach and Salpeter 1971;
Hollenbach, Werner and Salpeter 1971). More recently, Biham et al. (2001),
and Green et al. (2001) have computed $H_2$ production rate by physisorption. 
It was found that significant production is possible in cooler ($\sim 10-25$K) clouds
(see, also, Stantcheva, Shematovich \& Herbst 2002; Rae et al. 2003; Lipshtat et al. 2004).
Subsequently, Cazaux \& Tielens (2002, 2004) used both physisorption 
and chemisorption, especially to demonstrate that $H_2$ production is possible
at high temperatures ($\sim 200-400$K), as well. In these works, the H atoms 
combine together on the surface of the grains to form $H_2$ and then they are desorbed into the 
gas phase to react with other atoms. Whereas the investigations have been made to generally 
understand the rate of such reactions in presence of constant accretion rate of H onto the 
grain surface, and perhaps with a single type of grains, one requires to investigate 
this problem afresh in a realistic situation of collapse of dense proto-stellar clouds where accretion rates
may vary in the presence of a grain size distribution.

In the first paper of this series, we investigate precisely this problem. For concreteness, 
we concentrate on cool, dense interstellar clouds. We  assume typical stationary 
models of proto-stellar collapse of these clouds and compute the formation of $H_2$ 
molecules as the flow collapses. For number and size distributions of 
grains, we use the standard models of the grains (suitable for dense clouds) present in the literature. 
As far as the grain-surface chemistry goes, we compute using both the rate equation 
approach (e.g., Hasegawa, Herbst and Leung 1992) as well as the master equation 
(probabilistic) approach (e.g., Biham et al. 2001, Green et al. 2001).
The rate equation deals with the variation of the number density of the particles  
and probabilistic equation deals with the expectation value of the hydrogen number density. 
By comparing the $H_2$ formation in two methods, we conclude that the results of these 
approaches become identical at extremely small accretion rate 
as well as at large accretion rate $F_H>F_{H,c}$, where the 
exact value of the {\it critical accretion rate} $F_{H,c}$ depends on the grain size. (For definition of 
$F_H$, see, Eq. 9 below.) Typically, as we show in Sec. 2.4,  $F_{H,c} \sim 0.1-2$/s. For 
grains of smaller size, $F_{H,c}$ is larger, i.e., rate equations can be used if 
the accretion rate is sufficiently large. The first procedure is computationally faster and is valid
when the average number of each species on the grain is high. The second procedure
is slower and is useful when the average number of species is low. In a realistic collapse, 
the value of $F_{H,c}$ depends on the effective area of the grains, i.e., the sum of 
all the grain surfaces.  After following such self-consistent procedures, we conclude that 
a significant $H_2$ is formed in the cloud during the collapse. Since $H_2$ is a precursor of 
more complex bio-molecules, we believe that future works on bio-molecule formation
based on our present results would be more reliable.

The plan of the present paper is the following: In \S 2, we discuss the grain size 
distribution in a cloud. We also discuss the rate and the master equations. We show that the 
contribution of small grains towards the grain chemistry is more significant since 
their number density is much higher. We first compute recombination efficiency $2 R_{H2}/F_H$
(Biham et al. 2001) as a function of temperature for two different activation barrier energies
in order to select the suitable temperature range in which $H_2$ formation could 
become significant in the cold cloud condition. We compare the results $H+H \rightarrow H_2$ 
on grains using both the rate equation and master equation approaches as a 
function of accretion rate of $H$ and show that at high and very low rates, the results from these
two approaches merge. In \S 3.1, we first discuss our results for the case when the cloud is static, i.e.,
each shell starting with the same initial condition. We consider the cloud of high and low mass and of
two different temperatures.  We then show the variations of $H$ and $H_2$ as functions of the radial
co-ordinate. Since the cloud parameters are not very accurately known, we carry out the 
procedure for the cloud temperatures $T=10$K and $T=11$K respectively, as appropriate 
for a cold, dense cloud. In \S 3.2, we consider collapse of spherical shells. Here, the number 
densities are high, and thus the formation of $H_2$ on the grains and the subsequent desorption 
of $H_2$ to the gas are significant. We chose two types of grains such as olivine 
(at $10$ and $12$K) and amorphous carbons ($20$ and $25$K respectively) and use 
corresponding activation barrier energies. Finally, in \S 4, we make concluding remarks.

\section{Cloud and Grain Properties}

\subsection{Cloud Properties}

A generic interstellar cloud may have several distinct regions 
(e.g., van Dishoeck et al. 1993). Diffused clouds have typical number density
$\sim 100-800$cm$^{-3}$ and temperature $T \sim 30-80$K and a mass could vary from $1M_\odot$ to $100M_\odot$.
Translucent clouds are in the similar mass range, but have number density $\sim 500-5000$cm$^{-3}$ 
with a bit lower temperature ($15-50$K). Cold dark clouds have temperatures around $10-25$K, but the 
number densities are much higher ($10-10^4$cm$^{-3}$) and the corresponding masses can vary from
$0.3-10M_\odot$ at the core to $10-10^3M_\odot$ at  the cloud region. In terms of extinction parameter
$R_v$, it appears that a diffused cloud has $R_v \sim 3.1$, an intermediate dense region
has $R_v \sim 4.0$ and a dense cloud has $R_v\sim 5.5$ (Weingartner \& Draine, 2001a).
In contrast, the giant molecular clouds will have densities around $300$cm$^{-3}$ at the outer 
complex, $10^2-10^4$cm$^{-3}$ at the clouds, $10^4-10^7$cm$^{-3}$ in the warm 
regions while $10^7-10^9$cm$^{-3}$ in the hot cores. The typical size of a molecular 
cloud is around a parsec or so with an average lifetime $\sim 10 - 20$ Myr. 
In the outer regions, the cloud may pass through an isothermal phase  
where the temperature is constant. Heat generated in this phase is radiated away through 
the optically thin region. As the cloud collapses, the radiation is trapped due to the high 
optical depth and the gas becomes more adiabatic (Hartmann 1998). Though our numerical approach is
capable of handling temperature variation inside a cloud, in the absence of a suitable
and satisfactory distribution, in the present work,
we shall consider the clouds of constant temperatures for the sake of simplicity. 

\subsection{Size Distribution of Grains}

A small fraction (in terms of the number density) of the interstellar cloud is
in the form of dusts and grains. These grains play an important role in the formation of stars and  planetary 
systems and  in our context, causing chemical evolution of molecules. The degree of importance of the dust
depends on the ratio of the number density of grains to gas atoms or molecules at each radius.
Here we discuss two models of grains which we follow:

\noindent (a) MRN model:

The simplest model which can reproduce functional dependence of the extinction
with wavelength is due to Mathis, Rumpl and Nordsieck (1977). According to this so-called MRN model,
the number density of grains having size between $a$ to $a + da$ is given by, 
$$
dn_{gr} = Cn_Ha^{-3.5}da \hskip .1cm,\hskip .4cm     a_{min}< a < a_{max}\hskip .1cm, 
\eqno{(1)}
$$ 
where, n$_H$ is the number density of hydrogen, $a$ is the grain-radius in $cm$. This relation is 
strictly valid between the minimum and the maximum size of the grains $a_{min}$ = $50$ A$^o$ 
and $a_{max}$ = 2500 A$^o$. The grain constant, taken from Draine and Lee (1984) is given by $C= 10^{-25.13}$.

\noindent (b) Weingartner and Draine (WD) model: 

Weingartner and Draine (2001b, hereafter WD01) revised the MRN distribution for 
the entire range of grain size which is applicable to Milky way, LMC and SMC. This revision was needed 
to incorporate the real interstellar environment through which starlight passes. 
The parameters include reddening and carbon abundances. For instance, 
when the carbon abundance is negligible, this distribution becomes close to the MRN distribution. 
This distribution is valid for grains with $a>3.5$ Angstrom. If we take the weighted surface
area (i.e., surface area of the grain $\propto a^2$) multiplied by the number-density, that would
give the effective surface area per unit volume contributed by the grains of a given size $a$. In Fig. 1, we present this
plot for the parameters suitable for dense interstellar cloud 
($R_v^b = 5.5 $ and Case A from Table 1 of WD) used in the WD distribution.
It is clear that the small sized grains contribute the most because of their large number density. It is also clear that
there are three  sizes of grains which are important: (a) smallest one with $a \sim 10$ Angstrom (Type 1),
(b) the intermediate one with $a \sim 35$ Angstrom (Type 2) and (c) large grains with $a \sim 200$ 
Angstrom (Type 3). The actual values, marked by arrows have been computed from the humps in the WD distribution.
For simplicity of computation of the accretion rate on each type of grain and their 
catalytic effect, we assume that only these three types of grains exist. Their number densities
are estimated self-consistently from area under the curve given by WD distribution. See, Acharyya, Chakrabarti
\& Chakrabarti (2002, 2004) for application of this considerations.

\begin {figure}
\centerline{
\vbox{
\vskip -3.0cm
\psfig{figure=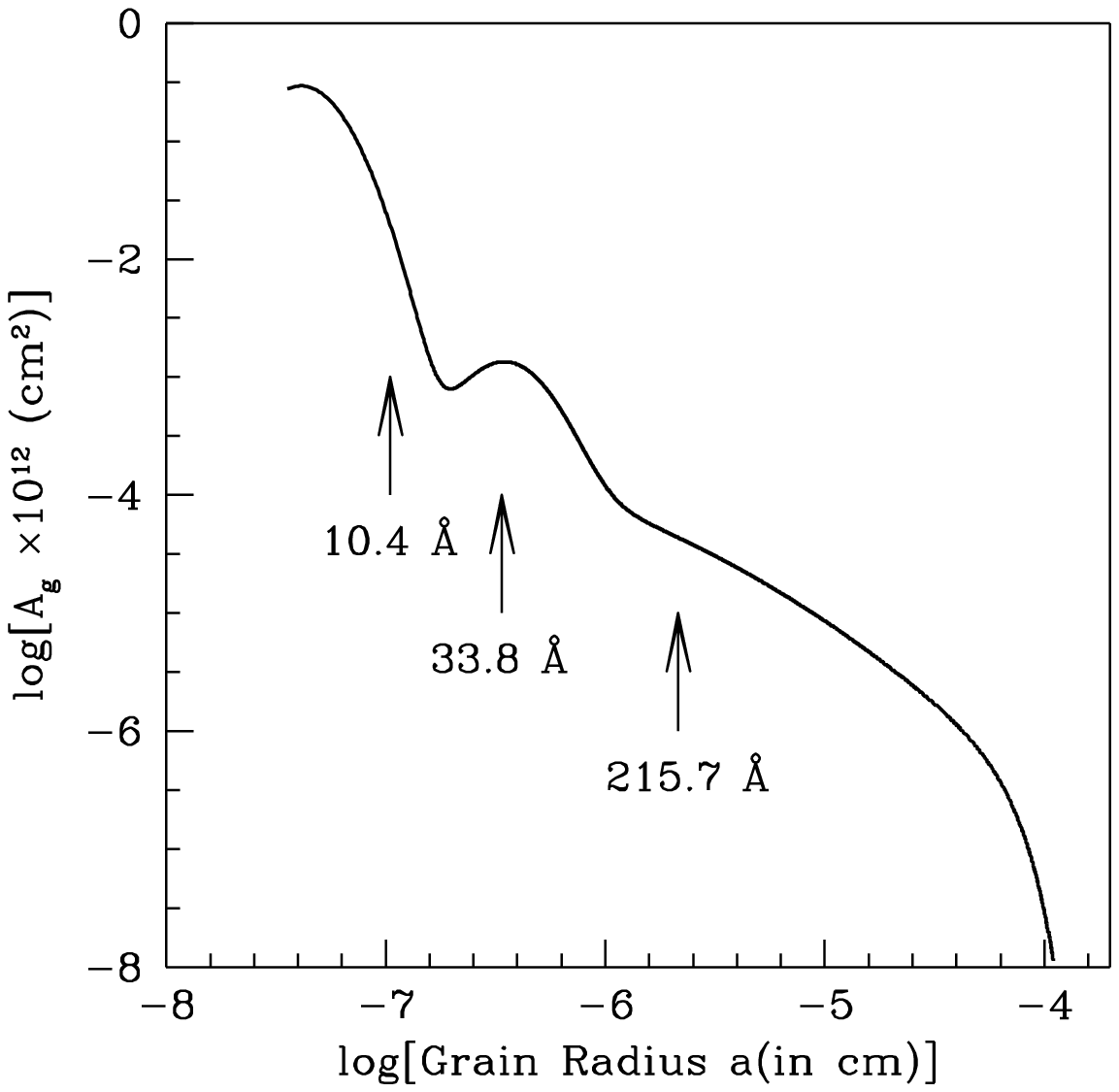,height=10truecm,width=14truecm}}}
\vspace{0.0cm}
\noindent{\small {\bf Fig. 1:}
Variation of the effective grain area $A_g$ as a function of the grain
radius indicating that smaller grains have the highest contribution. Instead of choosing
continuous distribution, three types of `average' grains have been assumed having
radii $10.4$ Angstrom (type 1), $33.8$ Angstrom (type 2) and $215.7$ Angstrom (type
3) respectively which have the same total number of grains. The arrows are drawn at
humps of the Weingartner \& Draine (2001b) distribution function. }
\end{figure}

\subsection{Equations governing the formation of molecular hydrogen}

\noindent{(a) Rate equation method}
 
Where the grain has large number of reactant atoms or molecules, it is convenient to use
this method. Here one deals with the average number of reactants.

Let us consider $n_H$ to be the number of H atoms on a grain at time $t$ and let $n_{H2}$ be the
number of $H_2$ molecules at that instant, then the following equation gives the rate at 
which number density of H changes:
$$
\frac {dn_H}{dt}=F_H - W_H n_H -2 (A_H/S)  n_H^2. 
\eqno{(2a)}
$$
Here, $F_H$  is the accretion rate of H which increases the number of H on a grain by sticking to it. 
$W_H$ is the desorption co-efficient of hydrogen =$\nu\exp(-E_1/k_bT)$, where, $E_1$ is 
the activation barrier energy  for desorption of H atom, $k_b$ is the Boltzmann's constant 
and $T$ is the temperature of the grain, assumed to be the same as the gas. Since in our case 
the number density is high $\sim 10^4$ cm$^{-3}$ and above, such an assumption is
justified. The second 
term causes a reduction of the number of hydrogen atoms on the grain, hence the minus sign. 
On the grain surface, mainly due to diffusive processes, two $H$ atoms combine to form a 
single $H_2$ molecule. $A_H= \nu\exp(-E_0/k_bT)$, the hopping rate, gives the probability of this to 
happen. Here, $\nu$ is the vibrational frequency,
$$
\nu= \frac{2 s E_d}{\pi^2 m_H}
\eqno{(2b)}
$$
where, $s \sim 10^{14}$ is surface density of sites on a grain (Biham, 2001), $m_H$ is the mass of the H atom and 
$E_d$ is the binding energy. This is normally taken to be $10^{12}$ - $10^{13} s^{-1}$.
$E_0$ is the activation  barrier energy for diffusion of H atom. This term also reduces the number of 
$H$ and hence the minus sign in the equation. The diffusion through tunneling has not been taken into 
account as it could be less important (Katz et al. 1999). $S$ is the number of sites per grain:  
$$
S=4 \pi r^2 s.
\eqno{(2c)}
$$
It may be relevant to discuss the nature of the third term of Eq. (2a). If the fraction $f_{gr}$ of 
grain-sites occupied by any of the species is very small ($f_{gr} <<1 $), then in order to form an $H_2$, 
the incoming $H$ has to hop, on an average $S$ times to combine with another $H$ located at an 
average distance of $S^{1/2}$ through a  well known `random-walk' process. This accounts for the 
factor $S$ in the third term which effectively reduces the diffusion rate. However, as
$f_{gr}$ becomes significant, any direction that the incoming $H$ hops to would be useful for
forming an $H_2$ and thus the factor may be reduced to $S^{1/2}$ or even less instead of $S$. In the 
present circumstance, $f_{gr}$ is indeed very small and we use the factor $S$ in the denominator as in
Biham et al. (2001) or Lipshtat et al. (2004).
 
The following equation gives the rate at which the number density of $H_2$ increases with time:
$$
\frac {dn_{H2}}{dt}=2 \mu (A_H/S) n_H^2 - W_{H2} n_{H2},
\eqno{(2d)}
$$
where, $W_{H2}$ is the desorption co-efficient of hydrogen molecule given by $\nu\exp(-E_2/k_bT) $,
$E_2$ is the activation barrier energy for desorption of $H_2$ molecule.
The parameter $\mu$ represents the fraction of $H_2$ molecule that remains on
the surface upon formation while (1-$\mu$) fraction is desorbed spontaneously due to the energy 
released in the recombination process. The $H_2$ production rate $R_{H2}$ in the gas due to grain is then given by,
$$
R_{H2}=(1-\mu) (A_H/S) n_H^2 + W_{H2}  n_{H2}. 
\eqno{(3)}
$$
However, the net production rate in the gas phase is obtained by inclusion of the breakdown of $H_2$ by 
cosmic rays, the rate of which is assumed to be $10^{-17}$ s$^{-1}$ (Millar et al. 1997). 
The values for energy barriers $E_0$, $E_1$, $E_2$ and $\mu$ are taken from Katz et al. (1999).
In our calculations, we used $E_0 = 24.7$ meV, $E_1 = 32.1$ meV, $E_2 = 27.1$ meV and $\mu = 0.33$ for olivine
and $E_0 = 44$ meV, $E_1 = 46.7$ meV, $E_2 = 46.7$ meV and $\mu = 0.413$ for amorphous carbon grains.  Because of the
difference in barrier energy, the recombination efficiencies (defined as $\eta = 2R_{H2}/F_H$, see, Biham et al. 2001)
are high at completely different temperature ranges. In Fig. 2(a-b), we show $\eta$ for (a) olivine and (b) amorphous
carbon computed using the rate equation. Here we also included the Langmuir \& Hinshelwood rejection term which rejects
the accretion of $H$ and $H_2$ in the occupied sites. This term essentially reduces the efficiency at lower temperature.
The curves are drawn for three accretion rates $F_H=10^{-8} s^{-1}$ (solid), $10^{-6} s^{-1}$ (dotted)
and $10^{-4} s^{-1}$ (dashed) respectively which are appropriate for the dense clouds considered in this paper. 
It is to be noted that for olivine (Fig. 2a), the useful temperature range is around $7-13$K and for carbon (Fig. 2b), 
this range is around $13-25$K. We therefore select the low cloud temperatures ($T=10$ and $12$K) for the first case
and relatively higher temperatures ($T=20$ and $25$K) for the second case. One could imagine having unusually high
accretion rates in dense clouds which might allow $H_2$ production at even higher temperatures where $\eta$ is still 
significant.

\begin {figure}
\centerline{
\vbox{
\vskip -3.0cm
\psfig{figure=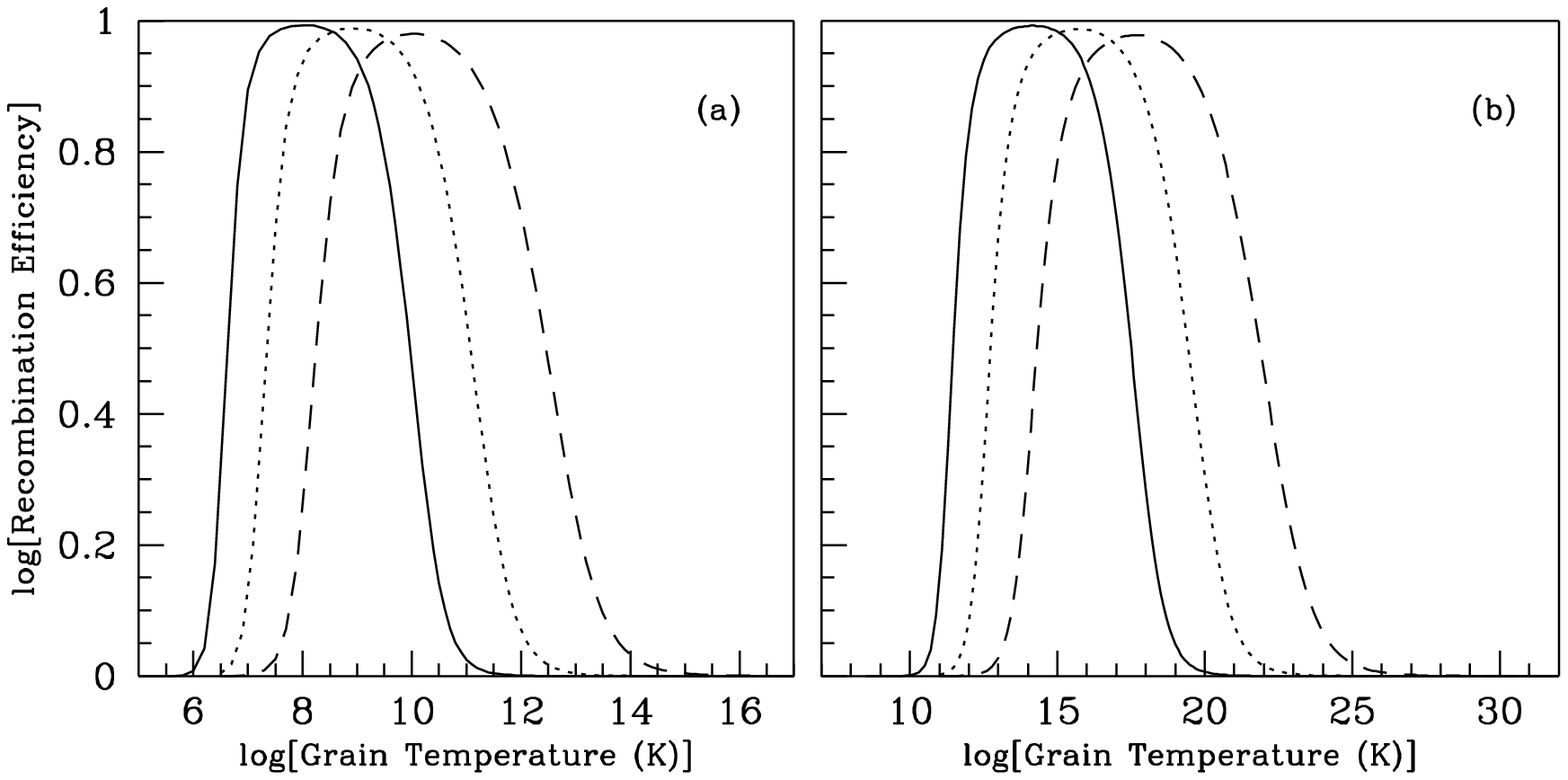,height=10truecm,width=14truecm}}}
\vspace{0.0cm}
\noindent{\small {\bf Fig. 2:} Variation of recombination efficiency $\eta$ as a function of the 
grain temperature in (a) olivine and (b) amorphous carbon respectively. The solid, dotted and dashed
curves are drawn for $F_H=10^{-8} s^{-1}$, $10^{-6} s^{-1}$ and $10^{-4} s^{-1}$ respectively.
}
\end{figure}

\noindent{(b) Master Equation Method}

It is convenient to use the master equation method (Biham et.al. 2001) to study the formation process 
when the number of species in  the grain is `small'. This process accounts for both the discrete 
nature of the hydrogen and the fluctuations and solves the problem probabilistically. For instance,  
its dynamical variables are the probability $P_H(N_H)$ that there are $N_H$ atoms on a grain
at a given time. One needs to study the time evolution of the probabilities through the same
type of equations as in Eq. 2(a-b), except that the expectation values $<N_H>$ and $<N_{H2}>$ 
are to be used instead of the average density. {\bf The equations are:
$$
\frac{d<N_H>}{dt} = F_H -  W_H<N_H> -2 (A_H/S) <N_H(N_H-1)> ,
\eqno{(4a)}
$$
and
$$
\frac{d<N_{H2}>}{dt} =\mu (A_H/S) <N_H (N_H-1)> -W_{H2}<N_{H2}> .
\eqno{(4b)}
$$
}
Since $P_H(N_H)$ represents the probability that there are $N_H$ hydrogen atoms on the grain,
by sum rule on probabilities: 
$$
\Sigma_{N_H=0}^{\infty}P_H(N_H)=1 .
\eqno{(5)}
$$
The time derivatives of these probabilities, ${\dot P}_H(N_H)$ are given by (Biham et al. 2001),
$$
{\dot P}_H(N_H)= F_H[P_H(N_H - 1)-P_H(N_H)] + W_H[(N_H + 1)P_H(N_H + 1)-N_HP_H(N_H)] +
$$
$$
~~~~~~~~~~~~~~~~~~~~~~~~~~~~~~~~~~~~ + (A_H/S)[(N_H + 2)(N_H + 1)P_H(N_H+2)-N_H(N_H - 1)P_H(N_H)].   
\eqno{(6a)}
$$
Similarly, the probability that there are $N_{H2}$ hydrogen molecules on the grain is 
given by $P_{H2}(N_{H2})$.  {\bf The time evolution of these probabilities is given by,
$$
{\dot P}_{H2}(N_{H2})= W_{H2}[(N_{H2} + 1)P_{H2}(N_{H2} + 1) - N_{H2}P_{H2}(N_{H2})]
$$
$$
+ \mu (A_H/S) <N_H (N_H-1)> [P_{H2}(N_{H2})-P_{H2}(N_{H2} - 1)] .
\eqno{(6b)}
$$
Note that the last term in Eq. 6b, does not exactly correspond to the last term in Eq. (13) 
of Biham et al. (2001). Perhaps there was a typographical error in the latter equation.

From these probabilities, one gets the expectation values for the number of H atoms on the grain as,
$$
<N_H> = \Sigma_{N_H=0}^{\infty}N_HP_H(N_H) ,
\eqno{(7a)}
$$ 
and the rate of formation of hydrogen molecules on the surface is,
$$
<N_{H2}> =  A_H \Sigma_{N_H=2}^\infty N_H (N_H-1) P_H (N_H).
\eqno{(7b)}
$$ 
The number of hydrogen molecules which are released back into the gas is given by,
$$
R_{H2}=(1-\mu) (A_H/S) <N_H (N_H-1)> + W_{H2} <N_{H2}>. 
\eqno{(8)}
$$
}

\subsection{Regime of Interest for Self-consistent Study}

Before we study the molecular hydrogen formation in a collapsing cloud,
we like to enquire if the rate equation or the master equation is to be used
for time evolution. For these, we require the accretion rate of $H$ falling on a given grain.
We compute three accretion rates for three different sizes of the grains.

As in the kinetic theory, the accretion rate is computed from the rate of $H$ which 
a grain `sees'. This is given by,
$$
F_H=\alpha \pi r^2 V n_h,
\eqno{(9)}
$$
where, $\alpha$ is the probability that a $H$ atom will stick to a grain,
$V$ is the root mean square velocity of the hydrogen which is given by,
$$
V=\sqrt{8 k_b T/\pi m_{p}},
\eqno{(10)}
$$
$n_h$ is the number density of hydrogen, $r$ is the mean radius of a 
grain $r_{grain}$. More accurately, it is the sum of the radii of the grain 
and the hydrogen atom. So we use, $r=r_{grain} + r_H$ ($r_H \sim 10^{-8}$ cm is the radius 
of a hydrogen atom). $\alpha$ may vary from 
$0.1$ to $1$.  We use $\alpha=0.5$ throughout the paper. $m_p$ is the mass of an hydrogen atom.
(van Dishoeck et al. 1993).

Figure 3 gives the time evolution of the variation of the number of $H$ and
$H_2$ on a grain surface at a given accretion rate of H for the small grains (Type 1) using both the
rate equation (long dashed) and  the master equation (solid curve). We choose 
the gas/grain temperature to be $10$ K. The accretion rate has been chosen to be $1.5\times 10^{-4}$.
Note that within about $10^4$s, both $H$ and $H_2$ saturate. This time is
negligible compared to the infall time. Thus, in our computation in future,
we assume that $H$ and $H_2$ saturate instantaneously.  

\begin {figure}
\vbox{
\vskip -3.0cm
\centerline{
\psfig{figure=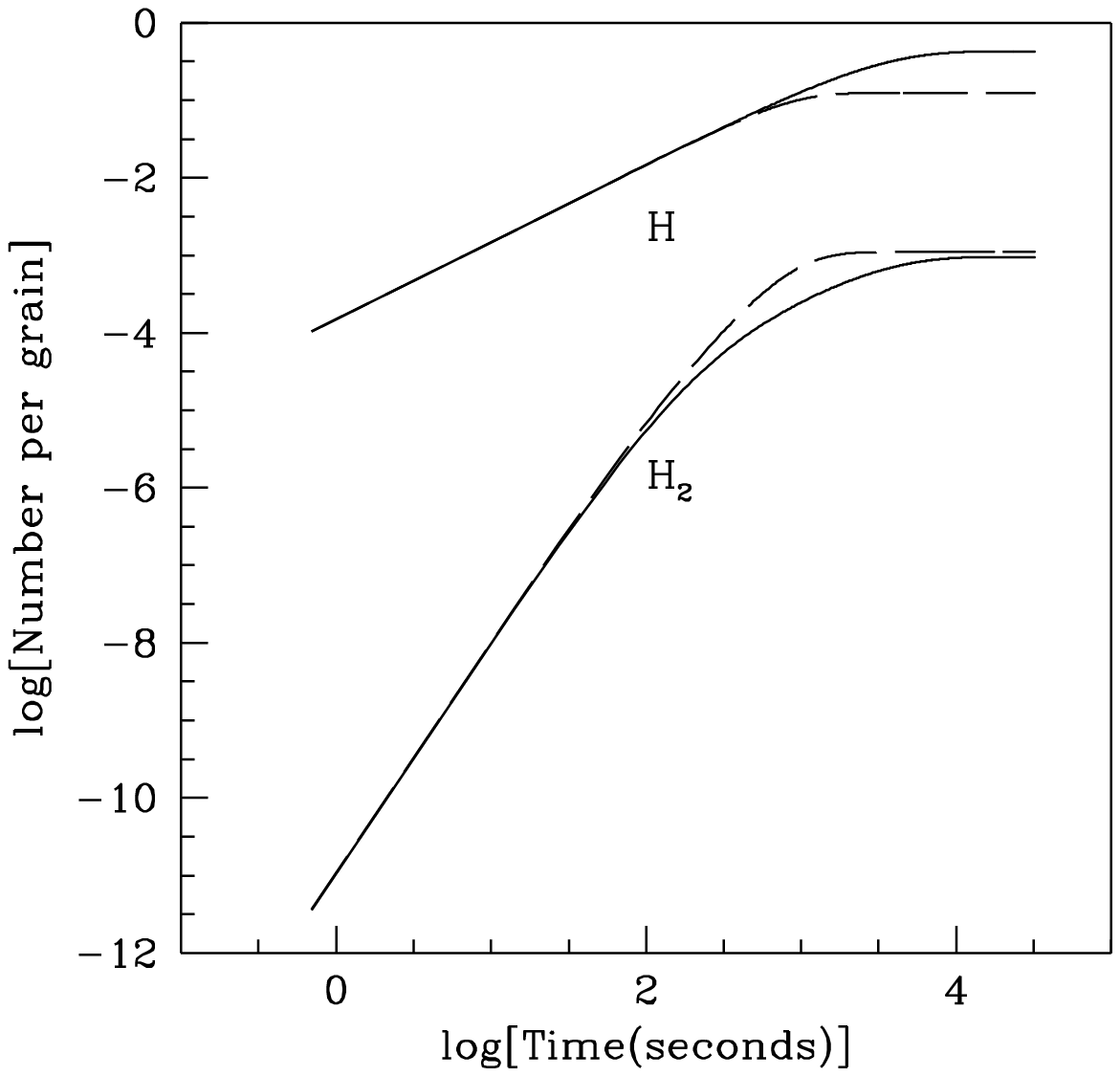,height=10truecm,width=14truecm}}}
\vspace{0.0cm}
\noindent{\small {\bf Fig. 3:}
Time evolution of the $H$ and $H_2$ numbers per grain using the average 
method, i.e., using the rate equation (long dashed curve) and the probabilistic method,
i.e., using the master equation (solid curve). The saturation takes place in less than a day.}
\end{figure}

In Figs. 4(a-b) we plot these saturation values for all three types of grains
as functions of accretion rate of H when the gas/grain temperature is $10$K. In Fig. 
4(a), we compare the number  of H for all three types of grains 
(marked) obtained from the rate (long dashed)  and from the master 
(solid curves) equations. Note that for extremely low value 
of the accretion rate, these two methods are close to each other. For sufficiently high rate, these two 
values merge. For lower rate, the results from the master equation
is independent of the grain size. If Fig. 4(b), we draw similar curves for the saturated 
values of $H_2$. The rate equation gives different saturation values for different types of grains (marked)
while the master equation gives virtually similar values.
Because in both the Figs. 4(a-b), the rate and master equations roughly merge
above the accretion rate of $F_{H,c} \sim 2$s$^{-1}$,  in our computation of $H_2$ formation during cloud collapse, we shall
assume that when the rate is  smaller than $2$s$^{-1}$, the master equation should be used, and
if it is larger than $2$s$^{-1}$, the rate equation should be used. This is therefore a
model-independent description and can be used even for time dependent work. We performed similar computations
in various temperatures and found the trend of the results to be similar.

\begin {figure}
\vbox{
\vskip -3.0cm
\centerline{
\hskip 1.0cm
\psfig{figure=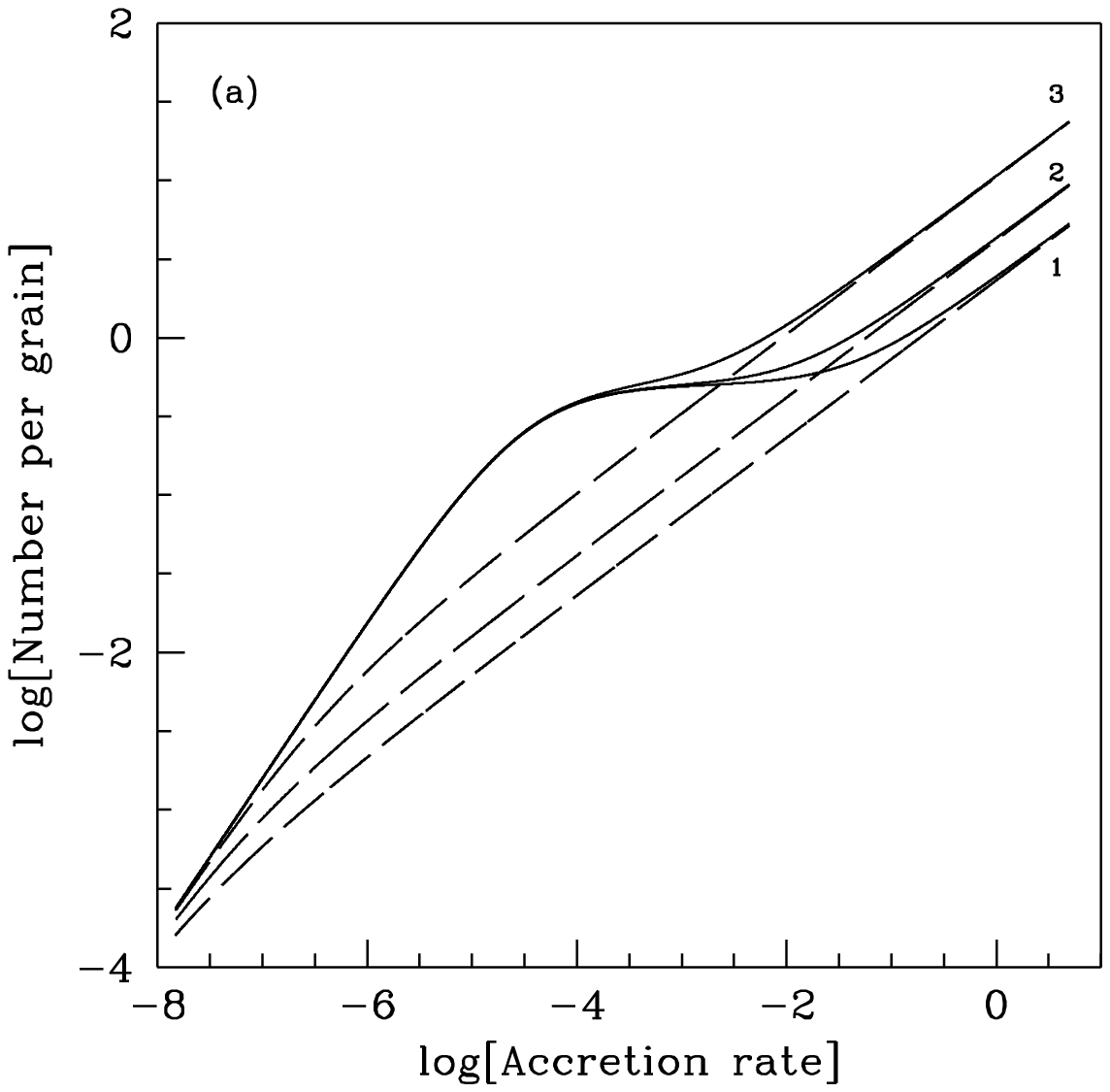,height=10truecm,width=12truecm}
\hskip -3.0cm
\psfig{figure=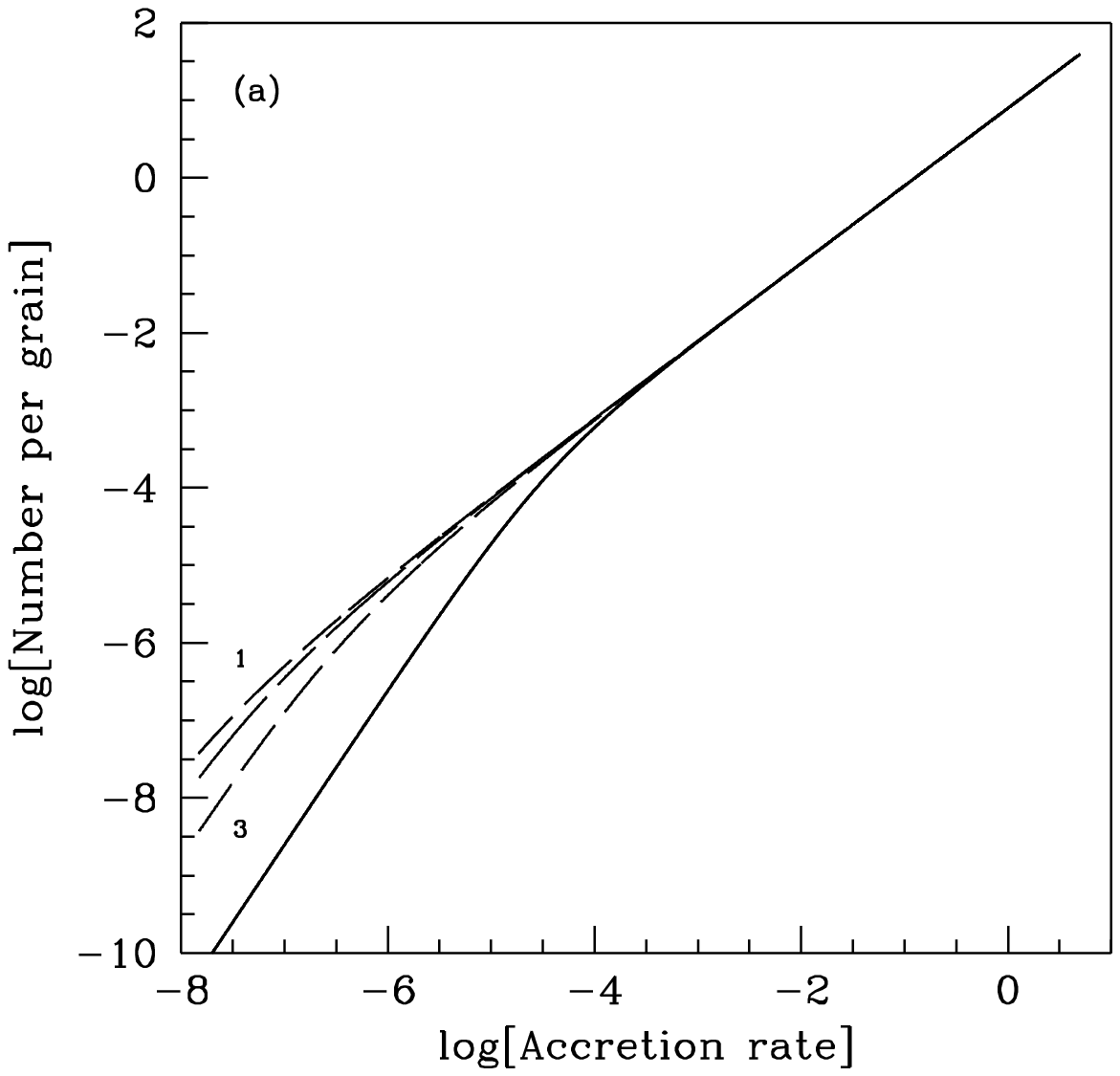,height=10truecm,width=12truecm}}}
\vspace{0.0cm}
\noindent{\small {\bf Fig. 4(a-b):}
Variation of the saturation values of the number of
(a) $H$ and (b) $H_2$ as a function of the accretion rate of hydrogen on different types of
grain surfaces (marked) when the gas/grain temperature is $10$K. Both the rate equation
(long dashed curves) and master equation (solid curves) have been used. In (a), 
the number of $H$ is the same from both the equations as very low and high rates.
In (b), only at high rates the rates are similar. This shows that above a critical
accretion rate of $\sim 1$s$^{-1}$ rate equations could be used for the production of $H_2$. }
\end{figure}

\section{Cloud Collapse and $H_2$ Formation}

So far, we presented the way the $H_2$ formation should be computed in presence of 
a size-distribution of grains.  We showed in particular that when the accretion rate is 
higher than certain value either the rate equation or the master equation
could be used, though it is computationally faster to use the rate equation approach.
In lower rates, only the master equation approach or the probabilistic approach should be used.
We now apply our results in two types of clouds.

\subsection{Evolution of $H_2$ Inside a Static Cloud}

In our first exercise, we assume an interstellar cloud which is static.
We start with a spherical molecular cloud having the outer radius at $R_{out}=1$pc.
The outer region of the cloud is assumed to be isothermal at $T_{out}$ where 
the gas is optically thin. $T_{out}$ is assumed to be a parameter. 
In this phase, we assume, $\rho\sim  r^{-2}$ (Shu, Adams and Lizano, 1987), where $r$ is the radial distance.
The accretion rate of $H$ on the grain surface will automatically increase as the density rises. 
In presence of rotation, centrifugal barrier is formed at $r=r_c$, where the 
centrifugal force balances gravity. At this stage, the cloud is expected to be 
disk-like and the opacity becomes high enough to trap radiations. Inside this 
region, $\rho \propto 1/r$. As a test of our code, we evolve the cloud  for 
ten million years to judge how far inside the cloud the production rate of $H$ is
significant.

Biham et al. (2001) pointed out, and we have also verified this to be true, that there
is indeed a narrow temperature range in which the recombination efficiency is the
highest. In the case of olivine the peak is at around $7-9$K and for amorphous
carbon the peak is at around $12-16$K. We chose temperatures at higher side of this
peak. In our model calculation we have taken two generic molecular clouds with $R_{out}=1 pc$
and $M=10M_\odot$ and $M=100M_\odot$. We assume the angular velocity of the outer edge of 
the cloud to be $10^{-16}$rad/sec so that $r_c=6.8 \times 10^{14}$cm 
and $r_c=6.8 \times 10^{13}$cm respectively in these two cases. 
With these masses, the densities of the gas at the outer edge are $\rho_{out}=0.162 \times 10^{-21}$gm cm$^{-3}$
and $\rho_{out}=0.162 \times 10^{-20}$gm cm$^{-3}$ respectively. Densities of the gas at $r_c$ are found out 
to be $\rho_c=0.324 \times 10^{-14}$gm cm$^{-3}$ and $\rho_c=0.262 \times 10^{-11}$gm cm$^{-3}$ respectively.
This density is the initial density of the collapsing shell.
A notional inflow velocity is  provided (which gives an indication of the evolution time
to be $t_{ev}\sim r/v$ by assuming a subsonic flow where the 
sound velocity is given by $a_s \sim (4kT/3m_p)^{1/2}$). In our runs, the 
velocity is always assumed to be less than this value. We 
choose $ \sim 10^4$ cm/sec for concreteness.  The activation barrier energies are chosen
to be as those of olivine.

The computational procedure involves in dividing the entire cloud into  $10^5$ shells. 
We compute the number density in each shell and the mean thermal velocity 
for temperature  $T=T_{out}$K. These enable us to get the accretion rate of $H$ on the grain surface.
The rate of desorption of $H_2$ which controls how much goes back into the gas at each
radius is computed. These are normalized by the number density of grains at each radius.
We continue our procedure for each shell. 

The results presented in this Section are for olivine grains.
In Figs. 5a, we present the mass fraction of $H$ and $H_2$ in the gas phase as a function 
of the radius of the cloud when $T_{out}=10$K when the mass of the cloud is $M=10M_\odot$ 
(dotted) and $M=100M_\odot$ (solid) respectively. Higher mass clouds produce significant $H_2$ at much
early stage of the evolution since the density is higher. In Fig. 5b, we present the
number densities of $H$ and $H_2$ on the grain surfaces. We remind here that the $H_2$ values
are actually the saturation expectation values at corresponding radii and $H$ values are 
the corresponding equilibrium values. $H_2$ goes back to the gas at the rate of $R_{H2}$ (Eq. 3).

\begin {figure}
\vbox{
\vskip -3.0cm
\centerline{
\hskip 1.0cm
\psfig{figure=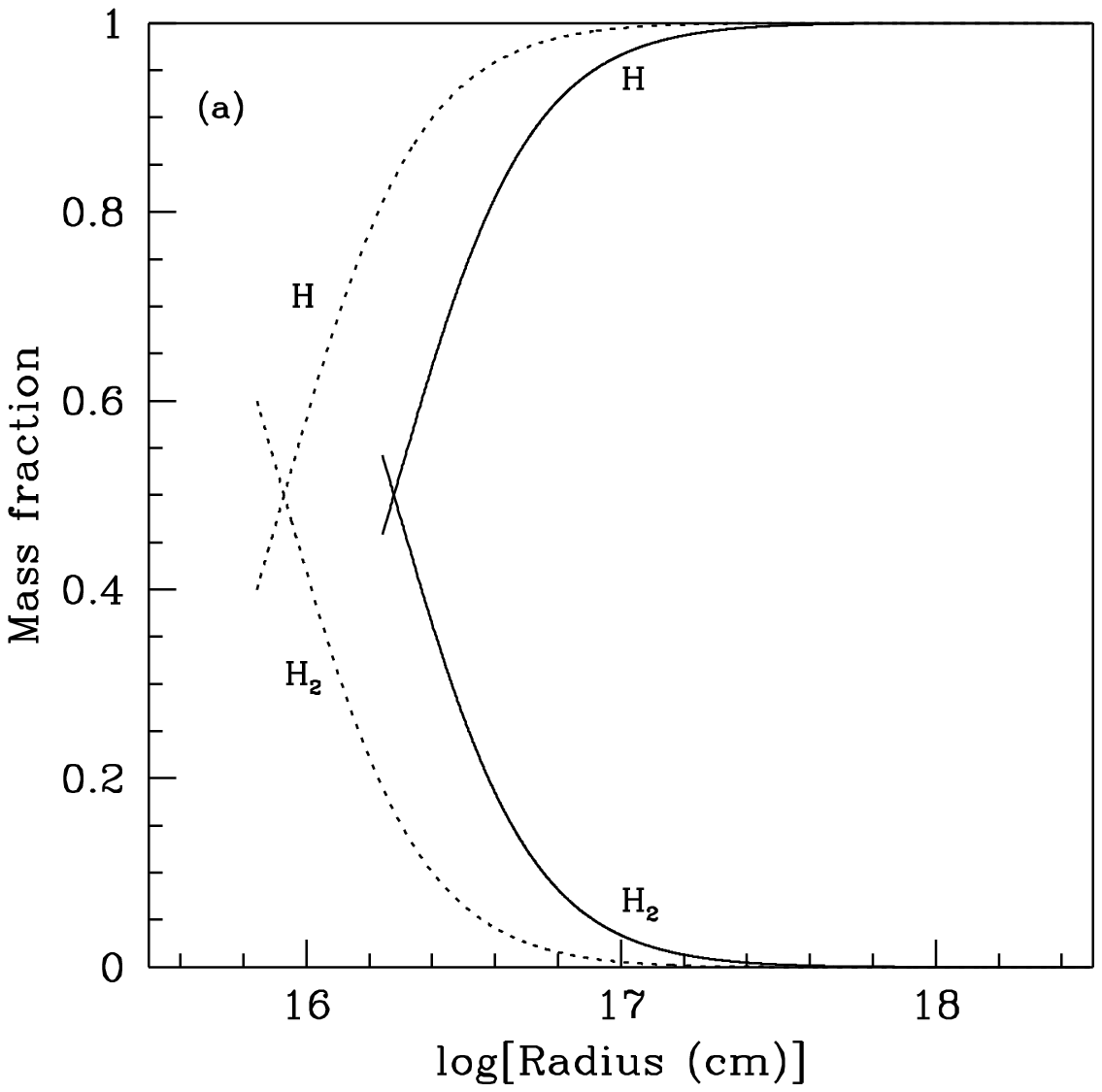,height=10truecm,width=12truecm}
\hskip -3.0cm
\psfig{figure=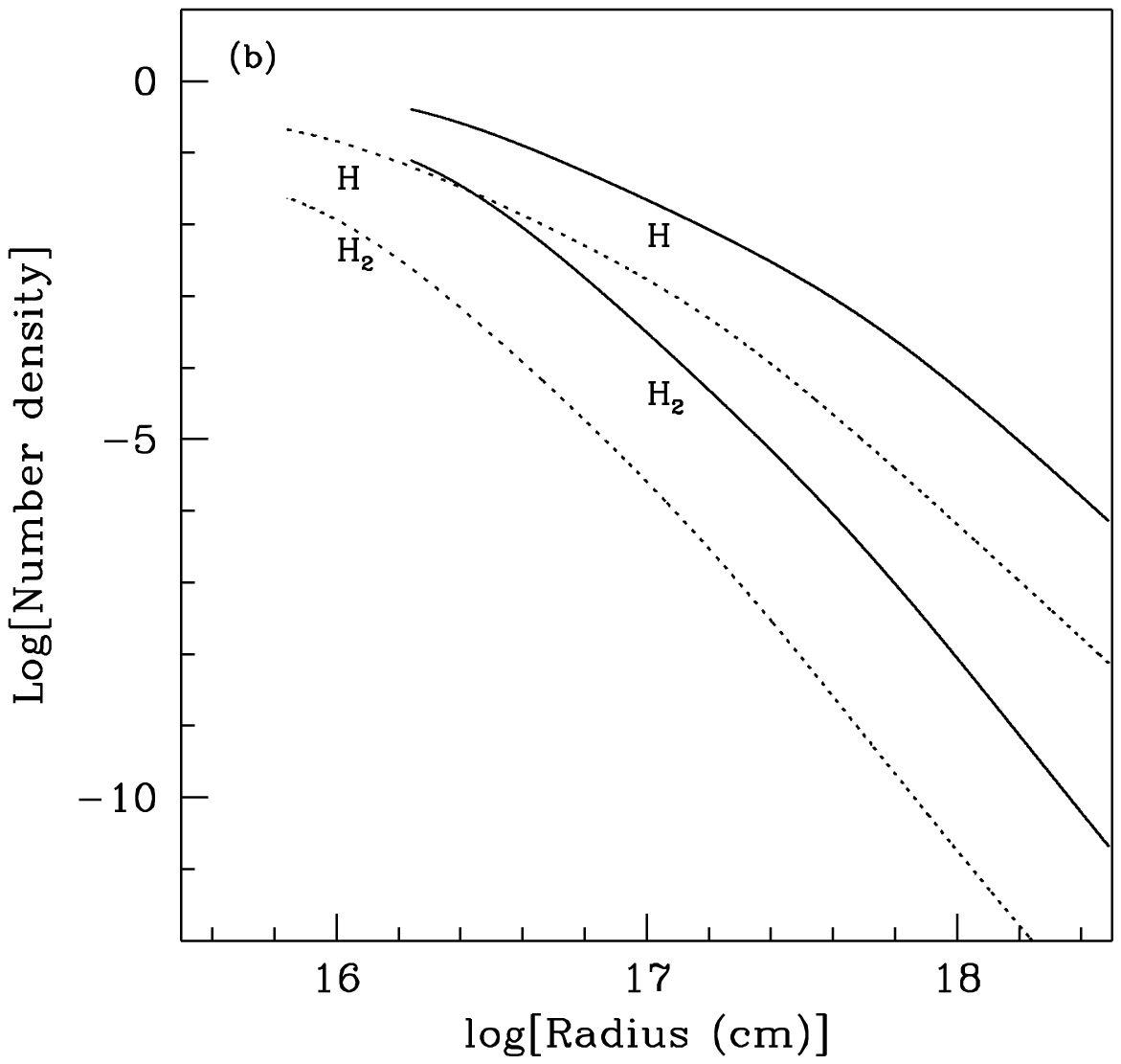,height=10truecm,width=12truecm}}}
\vspace{0.0cm}
\noindent{\small {\bf Fig. 5(a-b):}
Variation of (a) mass fractions of $H$ and $H_2$ in 
the gas phase and (b) number densities of $H$ and $H_2$ on a grain surface with radial distance
during the chemical evolution of a static cloud. 
$T=10$K is assumed. 
Dotted curves are for  $M=10M_\odot$ and the solid curves are for $M=100M_\odot$ respectively.  
For higher mass cloud the 
conversion to $H_2$ is faster because $H$ accretion rate on the grains is higher.}
\end{figure}

In Fig. 6(a-b) we present similar results at $T_{out}=11$K. 
The notations and styles are the same as before.
Generally speaking, the saturation values of $H$, $H_2$ on the grain surfaces 
increase with decreasing radius as before. Understandably, the production rate is 
higher when the cloud mass is higher because the accretion rate itself is high.
With the rise in temperature, $H$ is not significantly affected, but the production of $H_2$
in grains and therefore the desorption of $H_2$ into the gas are dropped. 
However, as one goes inside, the conversion of $H_2$ is higher because more $H$ was 
left over in the gas to increase the accretion rate on grains. As a result, the mass fraction of $H_2$ crossed 
$0.5$ mark at a larger radius. 

\begin {figure}
\vbox{
\vskip -3.0cm
\centerline{
\hskip 1.0cm
\psfig{figure=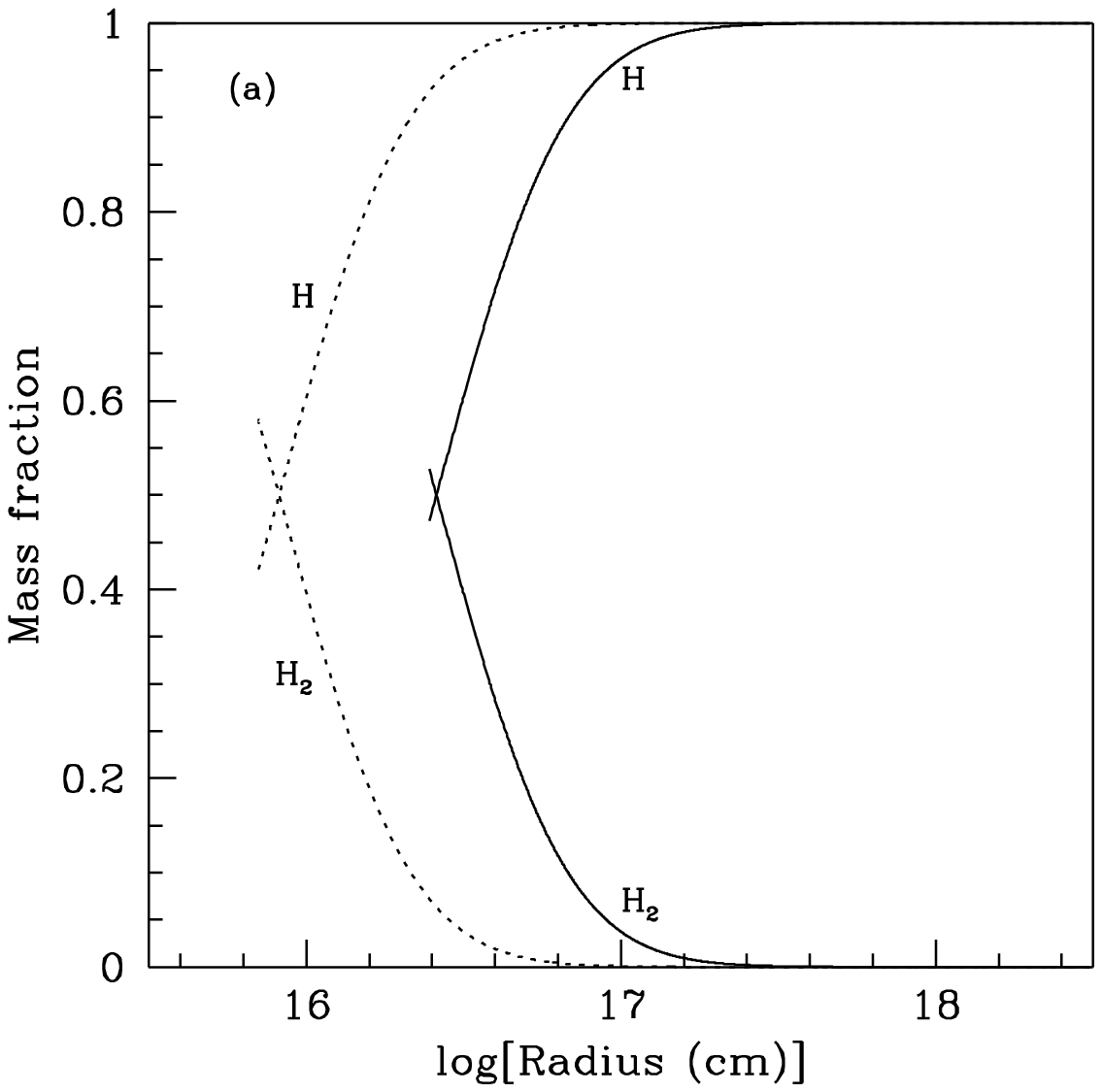,height=10truecm,width=12truecm}
\hskip -3.0cm
\psfig{figure=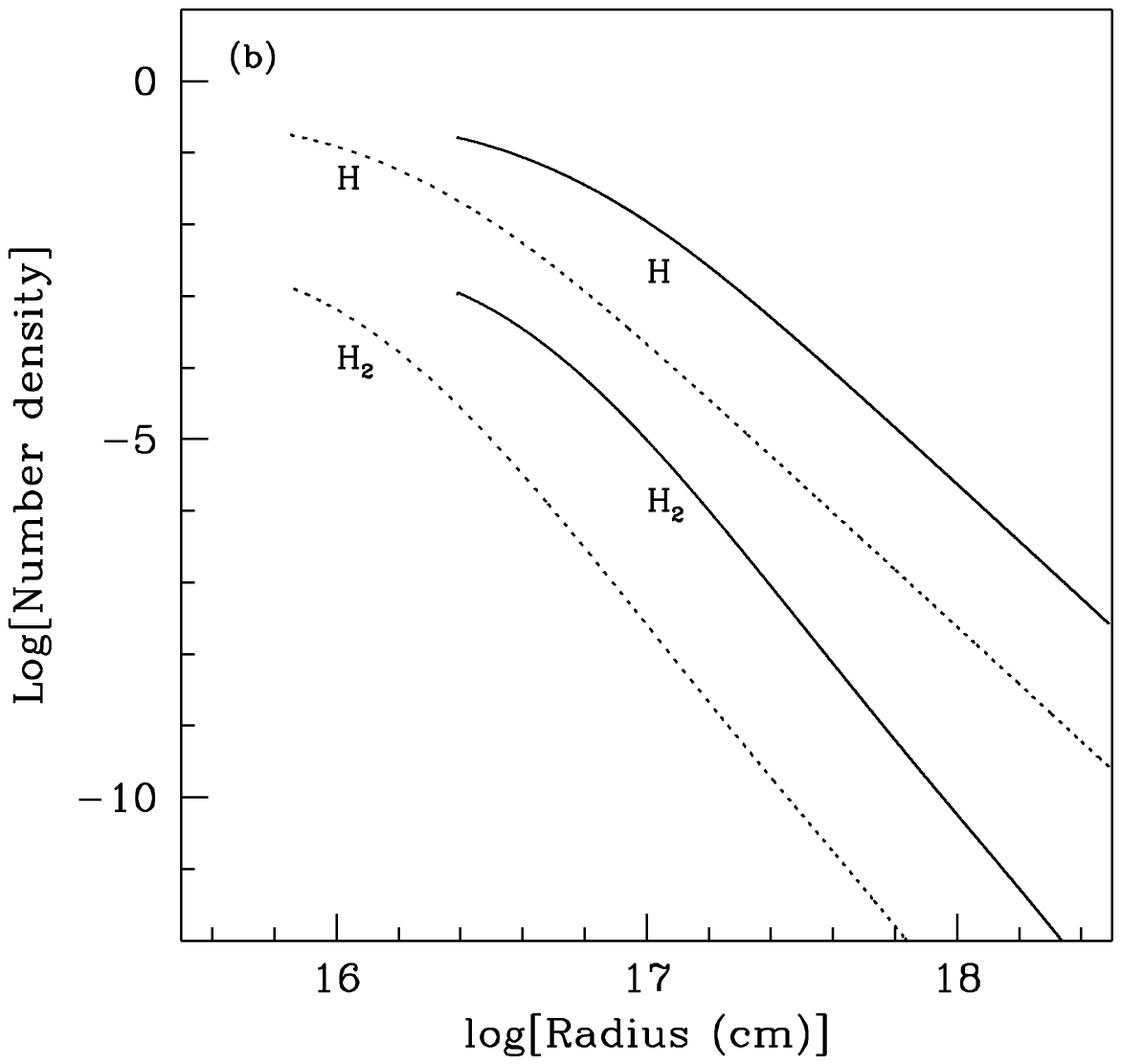,height=10truecm,width=12truecm}}}
\vspace{0.0cm}
\noindent{\small {\bf Fig. 6(a-b):}
Same as Figs. 5(a-b) but the temperature of the gas is $T=11$K.}
\end{figure}

\subsection{Collapsing Shell}

We now compute the variation of $H$ and $H_2$ with a somewhat different initial condition.
We start with a hollow spherical shell which is collapsing. Masses of the shell for two model runs 
are chosen to be $1M_\odot$ and $10M_\odot$ respectively. The initial densities are
$1.1 \times 10^{-20}$ gm cc$^{-1}$ and $1.1 \times 10^{-19}$ gm cc$^{-1}$ respectively. The shell thickness
is given by $0.0005$pc. In order to understand the effect of grain-types and the effect 
of temperature we choose two sets of activation energy barriers as presented in Sec. 2.3 
above. For each grain type, we choose two temperatures
for each of which we run two models with the mass of the shell taken to be $1M_\odot$ and $10M_\odot$ 
respectively. Our strategy is to compute the saturation value in this shell and let 
it collapse into the next shell freely. This becomes the initial condition of the 
next shell. The evolution proceeds from then on in that shell. The density increases 
in the same way discussed above while the mass of the shell remains fixed. For clarity, 
we find it convenient to plot the quantities with time of the collapse. The accretion rate 
of $H$ on the grains is very high in this model. The resulting number densities of 
$H$ and $H_2$ are also large. As a result, $H$ to $H_2$ conversion is very rapid
at early phases of the cloud collapse. 

\subsubsection{Olivine Shell}

First, we choose the activation energy barriers to be 
those of olivine (see Sec. 2.3). Figure 7a shows the evolution of $H$ and $H_2$ in the gas
phase when the cloud is of mass $M=1M_\odot$ (dotted) and $M=10M_\odot$ (solid) respectively. 
The temperature of the cloud has been chosen to be $T=10$K. Mass fraction of $H_2$ crosses $0.9$
at $t  \sim 7.81 \times 10^6 $yr and $t=4.7 \times 10^5 $yr respectively. In Fig. 7b, the number 
densities on the grain surfaces are plotted. Note that as the conversion to $H_2$ 
is rapid (Fig. 7a), the number density of $H$ decreases rapidly in the gas phase. 
At some point (around $T \sim 10^{5-6}$yr) the formation $H_2$ on grains is 
at a slower rate compared to the desorption rate) and hence the reserve of $H_2$
on grains goes down. On the other hand, the density of cloud itself increases as the 
collapse progresses which started increasing the accretion rate towards the end. This
increases the number density of the $H$ and $H_2$. Thus, in the collapsing shell model 
the number densities on the grains are lower at the intermediate times.

\begin {figure}
\vbox{
\vskip -2.0cm
\centerline{
\hskip 1.0cm
\psfig{figure=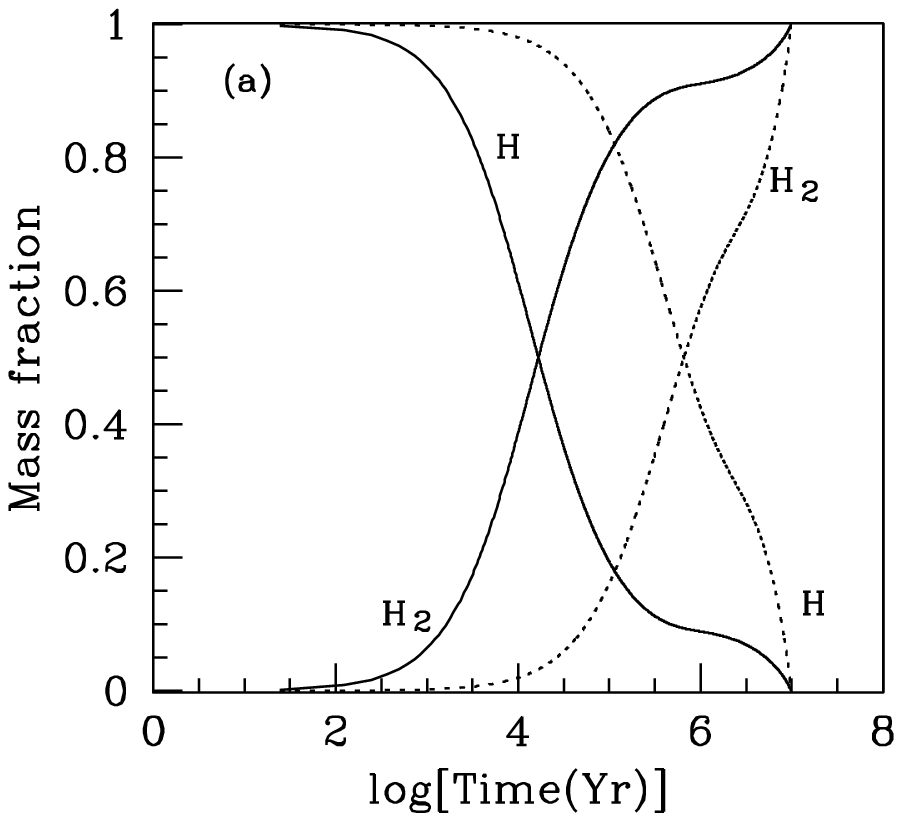,height=12truecm,width=13truecm}
\hskip -2.0cm
\psfig{figure=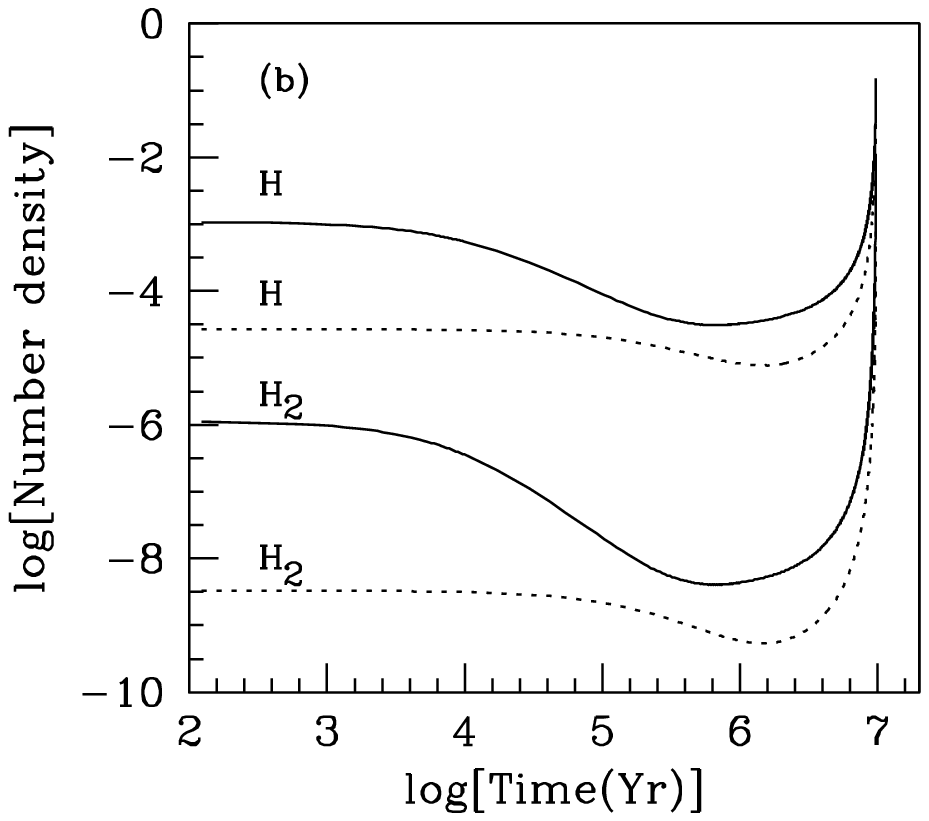,height=12truecm,width=13truecm}}}
\vspace{-2.0cm}
\noindent{\small {\bf Fig. 7(a-b):}
Variation of (a) mass fractions of $H$ and $H_2$ in    
the gas phase and (b) number densities of $H$ and $H_2$ on a grain surface
with time during the chemical evolution of a collapsing shell. Dotted curves 
are for  $M=1M_\odot$ and the solid curves are for $M=10M_\odot$ respectively.  
Mass fraction of $H_2$ reaches $\sim 0.9$ in a matter of $7.8 \times 10^6$yr and $4.7 \times 10^5$yr 
for low  and high mass shells respectively.}
\end{figure}

In Fig. 8(a-b) we repeat the same model (i.e., with the activation barrier energies as those of olivine)
but chose the cloud temperature to be slightly
higher at $T=12$K. The motivation is to show the sensitivity of the final outcome on temperature.
The behaviour is similar as in Fig. 7(a-b) except that the 
mass fraction of $H_2$  became $\sim 0.9$ at $t \sim 9.4 \times 10^6$Yr and $t\sim 8.1 \times 10^6$Yr
for $M=1M_\odot$ and $M=10M_\odot$ respectively. This is expected as the recombination efficiency is 
smaller (see, Fig. 2).

\begin {figure}
\vbox{
\vskip -2.0cm
\centerline{
\hskip 3.0cm
\psfig{figure=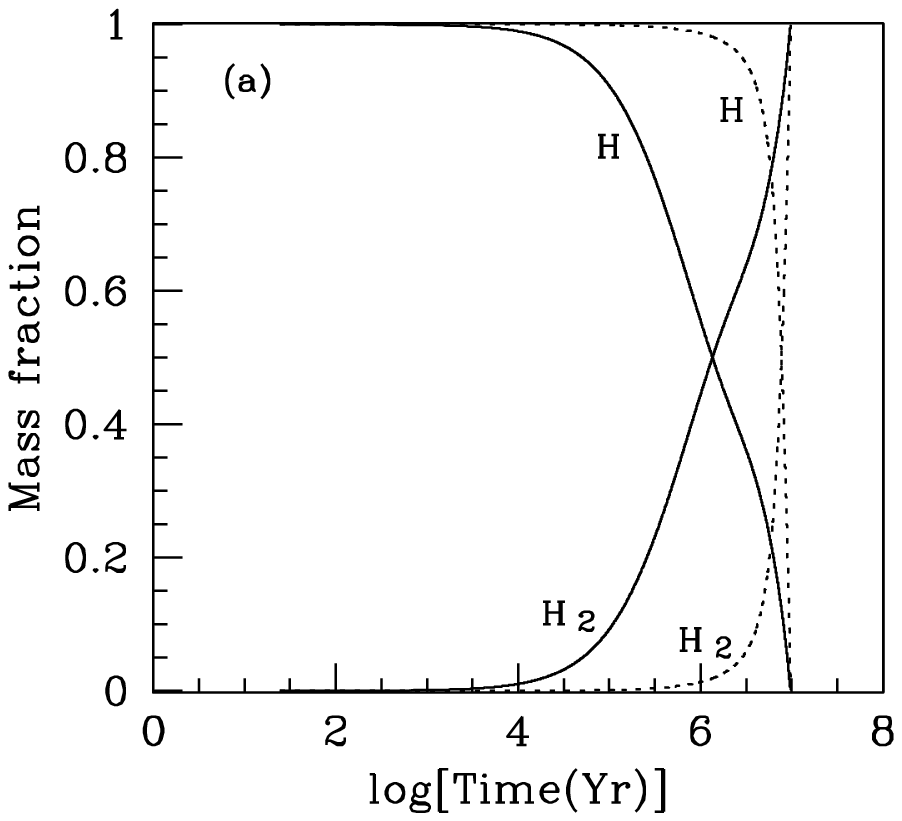,height=12truecm,width=13truecm}
\hskip -5.0cm
\psfig{figure=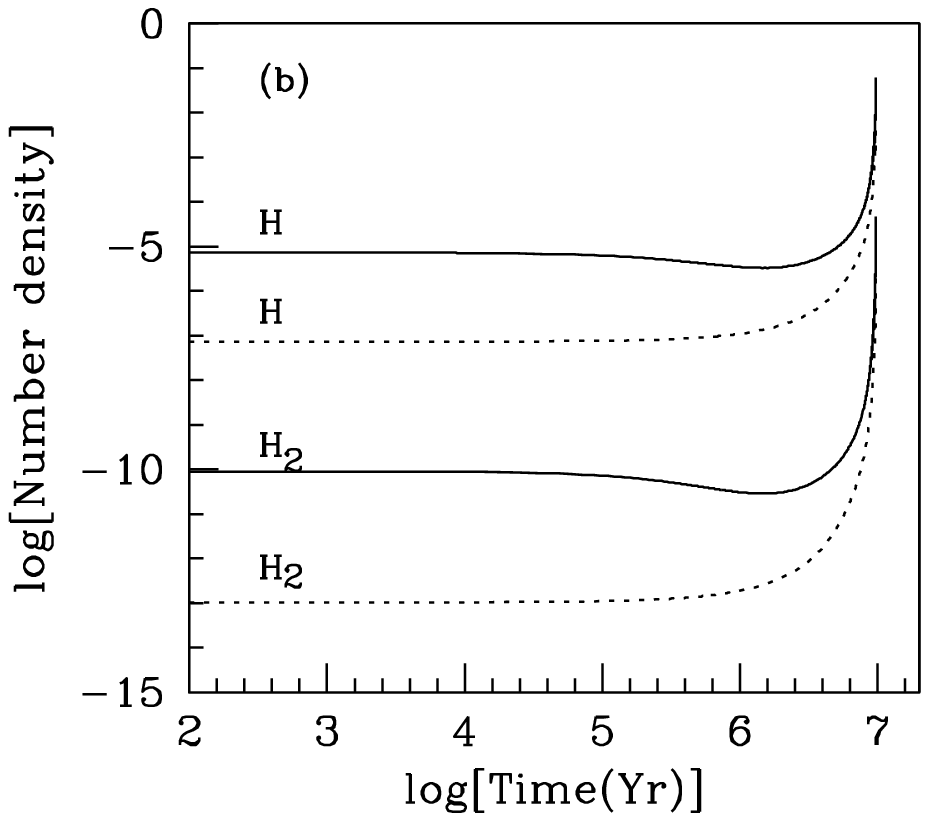,height=12truecm,width=13truecm}}}
\vspace{-2.0cm}
\noindent{\small {\bf Fig. 8(a-b):}
Same as in Fig. 7(a-b) but the temperature of the
cloud is assumed to be $12$K. Mass fraction of $\sim 0.9$ is achieved by $H_2$
in a matter of $9.4 \times 10^6$yr and $8.1 \times 10^6$yr for low and high 
mass shells respectively.}
\end{figure}

\subsubsection{Amorphous Carbon Shell}

We now assume the activation barrier energies chosen to be those of amorphous carbon grains.
In this case, it is possible to go to higher temperatures. We run at two temperatures,
$T=20$K and $T=25$K respectively for the shells of $M=1M_\odot$ and $M=10M_\odot$
respectively. In Figs. 9(a-b), we plot the results for $T=20$K. Here, the plot 
characteristics remained the same as before. The mass fraction of $H_2$ reaches $0.9$
at $t= 9.1 \times  10^6$ and $t=6.4\times 10^6M_\odot$ years for $M=1M_\odot$ and $M=10M_\odot$
respectively. In Fig. 9b, we show the number densities of $H$ and $H_2$ on grains. 
Here too, there is a minimum at an intermediate time.
Similar Figures as above  for $T=25$K are plotted in Figs. 10(a-b) and the
mass fraction of $H_2$ reached $0.9$ at $t=9.5 \times 10^6$yr and $t=9.36\times 10^6$yr
respectively. Thus, in the case of higher activation energy, the sensitivity of
the final result on the shall mass is weaker.

\begin {figure}
\vbox{
\vskip -2.0cm
\centerline{
\hskip 3.0cm
\psfig{figure=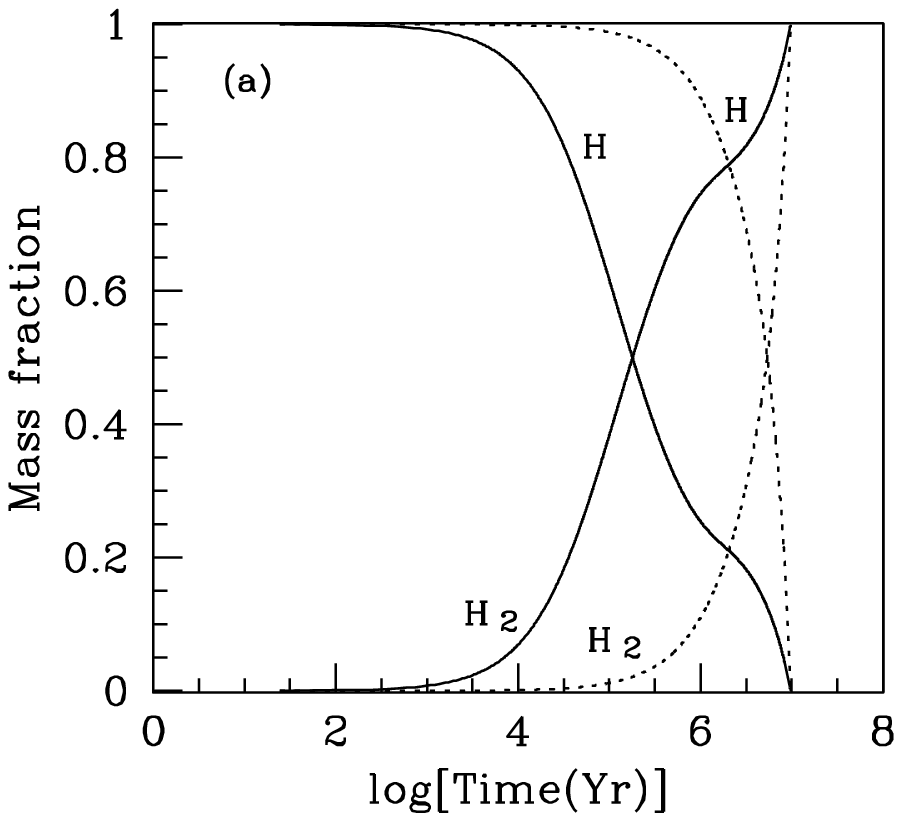,height=12truecm,width=13truecm}
\hskip -5.0cm
\psfig{figure=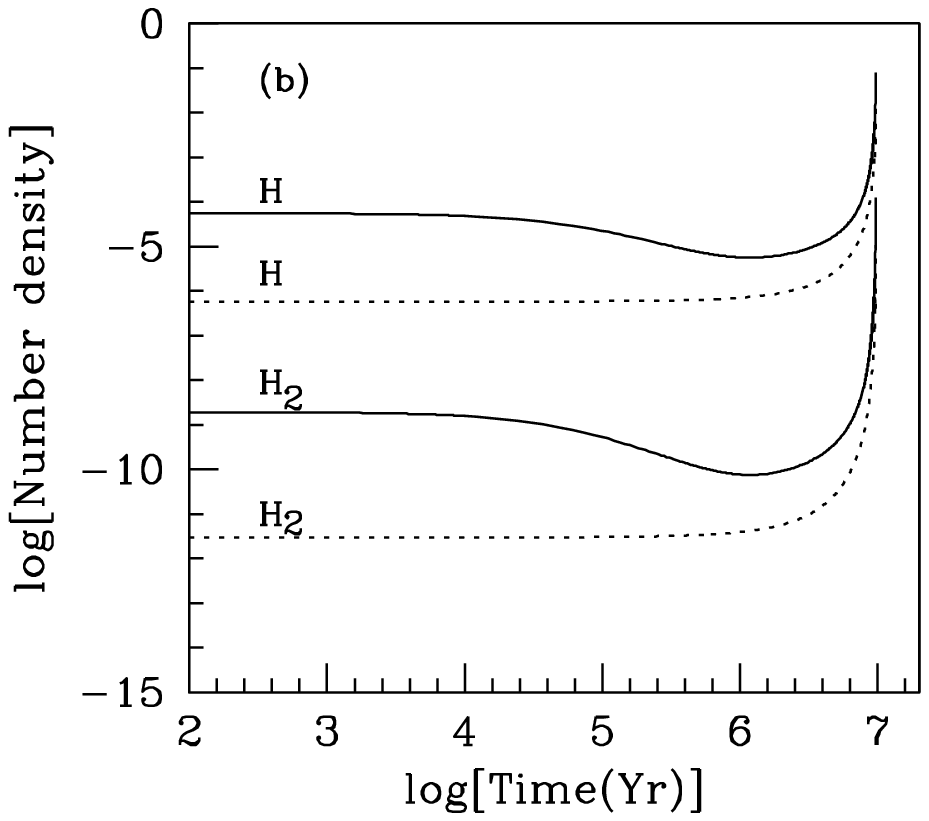,height=12truecm,width=13truecm}}}
\vspace{-2.0cm}
\noindent{\small {\bf Fig. 9(a-b):}
Same as in Fig. 7(a-b) but the activation barrier energies of the
cloud is assumed to be those of amorphous carbon. The temperature is chosen to be $T=20$K.
Mass fraction of $\sim 0.9$ is achieved by $H_2$ in a matter of $9.1\times 10^6$yr 
and $6.4\times 10^4$yr for low ($M=1M_\odot$) and high ($M=10M_\odot$) mass shells respectively.}
\end{figure}

\begin {figure}
\vbox{
\vskip -2.0cm
\centerline{
\hskip 3.0cm
\psfig{figure=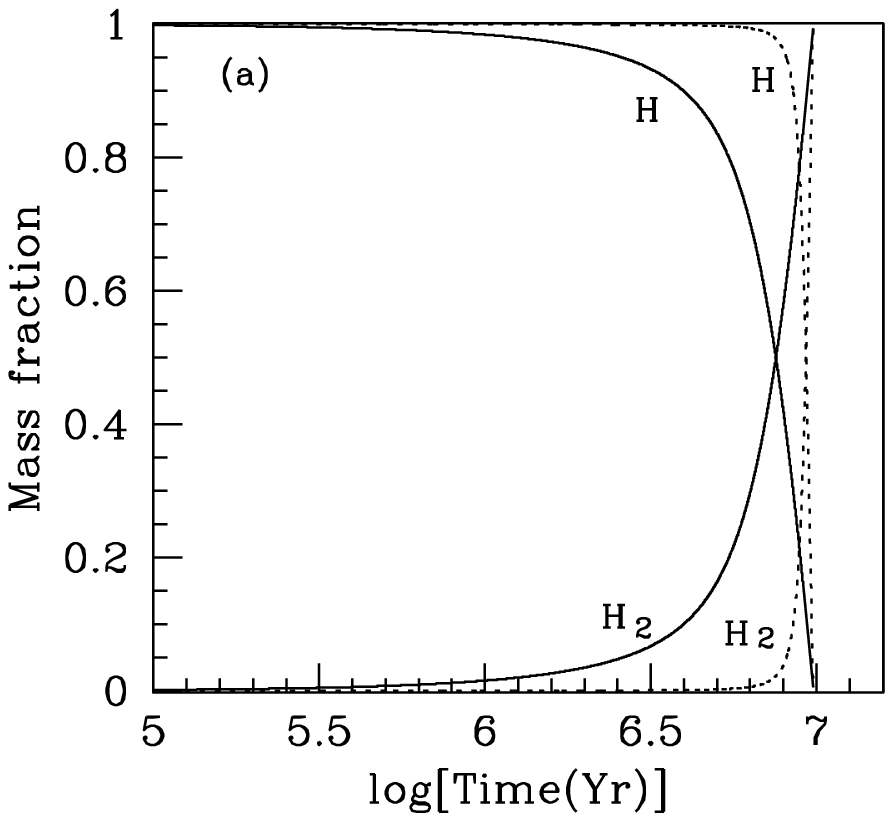,height=12truecm,width=13truecm}
\hskip -5.0cm
\psfig{figure=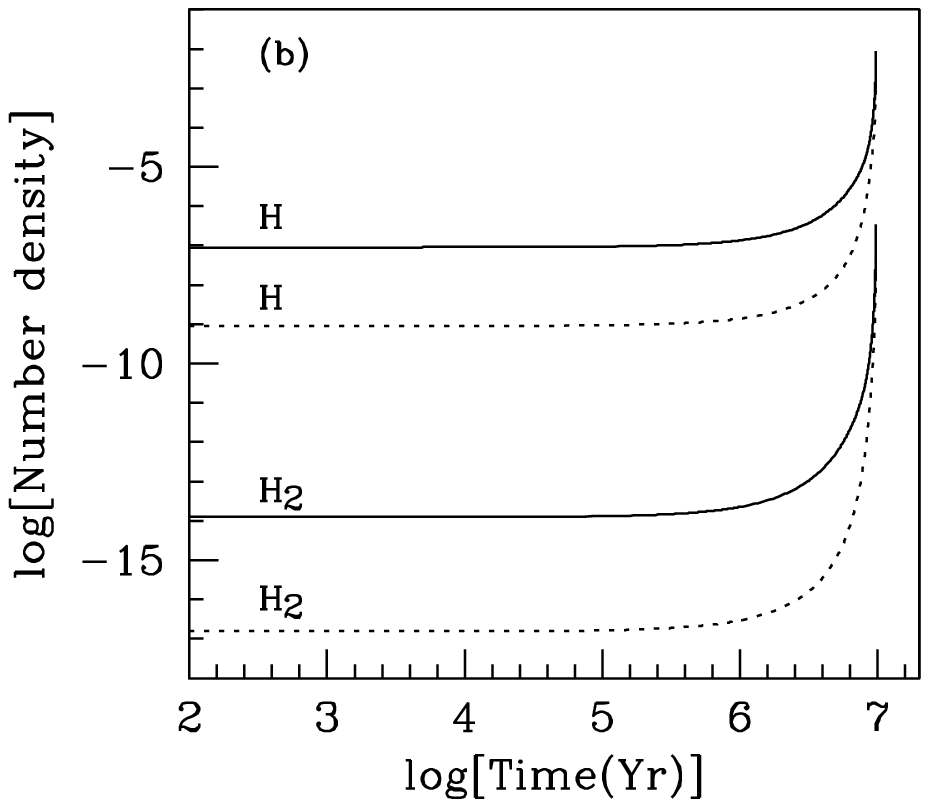,height=12truecm,width=13truecm}}}
\vspace{-2.0cm}
\noindent{\small {\bf Fig. 10(a-b):}
Same as in Fig. 8(a-b) but the temperature of the cloud is assumed to be $25$K. 
Mass fraction of $\sim 0.9$ is achieved by $H_2$
in a matter of $9.5 \times 10^6$yr and $9.36\times 10^6$yr for low ($M=1M_\odot$) and high 
($M=10M_\odot$) mass shells respectively.}
\end{figure}

In this context, we wish to mention that while computing the formation of $H_2$ using a 
constant surface reaction rate supplied by UMIST database (Millar et al. 1997)
it was observed that most of the $H$ is converted to $H_2$ at a somewhat later 
stage (Chakrabarti \& Chakrabarti, 2000), at least after $10^7$yr of the 
begining of the collapse. Our work with computation with up to date grain 
surface chemical processes indicates that  the conversion is perhaps faster. 
In future, we shall compute more complex molecules and would compare them with 
observed tabulated value (e.g. Herbst 1992 and van Dishoeck et al. 1993).

\section{Discussion and Concluding Remarks}

In this paper, we studied evolution of molecular hydrogen in the gas phase 
and on the grain surface in two toy models, namely, in (a) a static interstellar cloud and (b) the
collapsing phase of a spherical shell. In the later case, we chose 
two different types of activation barrier energies which correspond to olivine and 
amorphous carbon grains. Instead of using the complete
grain size distribution, we assumed three major grain sizes where there are humps in the
WD distribution. We compared the results from the rate and the master equation for each type of grains
and determined the applicability of these equations as a function of the accretion rate of $H$ on grains.
Using static and collapsing shell models with prescribed density, temperature and velocity distributions, 
we showed how gradually $H_2$ builds up in the grain and the gas. Earlier, using a constant and 
representative surface reaction rates given in the UMIST data base for $H$ to $H_2$ conversion,
it was shown  that the conversion process is slower (e.g. reaching mass fraction of $\sim 0.5$
after $t \sim 10^7$yr). However, here we find that ninety percent conversion to $H_2$ is achieved in a matter of 
$10^5-10^7$ years depending on cloud parameters such as activation energy and temperature of the grains,
and the mass of the cloud. In terms of spatial distribution, we find that most of the conversion takes place for
a radial distance $r>0.1$pc, i.e., at the outer shell of the molecular cloud.

In our work, we chose simple power-law density distribution inside an isothermal cloud.
Li and Draine (2001) considered the effect of radiations on the grains in diffused interstellar clouds 
and computed the grain temperature as functions of grain size distribution. They also
computed the resulting infrared spectra.
Recent numerical solutions of several workers such as Galli, Walmsley \& Goncalves 
(2002), Zucconi, Walmsley \& Galli (2001) have shown that dust temperature inside the dense cloud
could be as low  as $6$K and the temperatures of the gas and grain become roughly identical when the 
gas density $n \gsim$ few $\times  10^4$ cm$^{-3}$. Thus our assumption of the equality of the
grain and the dust temperatures may be justified since our number densities are of this order or higher. 
On the other hand, COBE/DIRBE results  (e.g., Lagache et al. 1998)
do not seem to support the existence of very cold dusts. We therefore chose the cloud temperature at $10$K and above.

Our work concentrated on the gravitational collapse of a dense cloud induced by Jean's 
instability. In a different context, Bergin et al. (2004) studied the evolution of the 
molecular hydrogen inside an extremely diffused cloud
the formation of which is induced by the passage of a shock wave. The temperatures used were very high 
(reaching several thousand) which is appropriate for such a cloud of very low extinction. In their work,
though the gas-grain interaction was simplified through by using a production rate of $H_2$ as 
given in Millar, Farquhar, Willacy (1997), they used destruction of $H_2$ through cosmic rays 
and UV radiation. They also studied the formation rate of CO molecule. 

In our paper, we have throughly dealt with the most basic surface chemistry,
namely the $H_2$ formation deep inside a cold, dense cloud. In future, we shall apply 
our procedure to include diffused clouds at high temperatures following above work and explore
the possibility of formation of more complex molecules and especially the 
formation of bio-molecules during the collapse phase.

\section*{Acknowledgments}
The authors thank the referee for very useful comments leading to improvements in the manuscript.
This work is supported in part by Indian Space Research Organization (ISRO) Grant from a RESPOND project.

{}

\end{document}